\documentclass[ twocolumn,aps,prd,   
               preprintnumbers,numbers,sort&compress,nofootinbib,
                            showpacs,superscriptaddress,
               colorlinks,
               linkcolor=blue,   
               citecolor=blue]{revtex4-1}   \newcommand{\exclude}[1]{}

\usepackage{graphicx,amsmath,amssymb,bm}
\DeclareMathOperator{\Tr}{Tr}
\usepackage{psfrag}
\usepackage{hyperref}
\usepackage{enumitem}
\usepackage{siunitx}

\newcommand{\beq}{\begin{equation}}
\newcommand{\eeq}{\end{equation}}
\newcommand{\be}{\begin{eqnarray}}
\newcommand{\ee}{\end{eqnarray}}
   
\def\dd{ \,\mathrm{d} }

\def\+{\dagger}
 \def\la{\langle}
 \def\ra{\rangle}

\begin{document}
\preprint{CALT-TH 2017-007}

\title{ Axion detection via Topological Casimir Effect  }

 \author{ChunJun \surname{Cao}}
\email{cjcao@caltech.edu}
\affiliation{Walter Burke Institute for Theoretical Physics California Institute of Technology, Pasadena, CA 91125, USA}
 
\author{Ariel \surname{Zhitnitsky}}
\email{arz@phas.ubc.ca}
\affiliation{Department of Physics and Astronomy, University of British Columbia, Vancouver, B.C. V6T 1Z1, Canada}

\begin{abstract}
We propose a new table-top experimental configuration for the direct detection of dark matter QCD axions  in the traditional open mass window $10^{-6} {\rm eV}\lesssim  m_a\lesssim   10^{-2} {\rm eV}$     using non-perturbative effects in a system with non-trivial spatial topology.
  Different from most experimental setups found in literature on direct dark matter axion detection, which relies on $\dot{\theta}$ or $\vec{\nabla}\theta$, we found that our system is in principle sensitive to a static $\theta\geq  10^{-14}$ and can also be used to set limit on the fundamental constant $\theta_{\rm QED}$ which becomes the fundamental observable parameter of the Maxwell system if some conditions are met. 
  Furthermore, the proposed experiments can probe entire open mass window  $10^{-6} {\rm eV}\lesssim  m_a\lesssim   10^{-2} {\rm eV}$   with the same design, which   should be contrasted with conventional cavity-type experiments  being  sensitive  to a specific axion mass.  
  Connection with Witten effect when the induced electric charge  $e'$ is proportional to $\theta$ and the magnetic monopole becomes the dyon with non-vanishing  $e'=-e \frac{\theta}{2\pi}$ is also discussed.
\end{abstract}


\maketitle

\baselineskip=15pt


\section{Introduction and Motivation}\label{introduction}
The leitmotiv of the present work is related to the fundamental parameter $\theta$ in the Maxwell Electrodynamics, as well as the axion field related to this parameter. The $\theta$ parameter was originally introduced in Quantum Chromodynamics (QCD) in the 70s. Although the term can be represented as a total derivative and does not change the equation of motion, it is known that this parameter is a fundamental physical parameter of the system on the non-perturabative level. In particular, $\theta\neq 0$ introduces $\cal{P}$ and $\cal{CP}$ violation in QCD, which is most well captured by the renowned strong $\cal{CP}$ problem.  A formal and deep reason of  why  the $\theta$   becomes  a real physical parameter 
is that the topological mapping  $\pi_3[SU(N)]=\mathbb{Z}$  in a 3+1 dimensional  world   is nontrivial. This rich  topological structure    leads to the generation of physically identical (but topologically distinct)   sectors which play the key  role  in non-perturbative  formulation of the  theory. 
 
  The strong $\cal{CP}$ problem in  QCD problem was resolved by promoting the fundamental parameter $\theta$ to a dynamical axion $\theta(x)$ field, see original papers \cite{axion1, axion2, axion3, KSVZ1, KSVZ2, DFSZ1, DFSZ2} and   review articles \cite{vanBibber:2006rb,Asztalos:2006kz,Raffelt:2006cw,Sikivie:2009fv,Rosenberg:2015kxa,Graham:2015ouw}. However, the axion has not yet been discovered 40 years after its initial formulation. Still, it remains one of the most interesting resolutions of the  strong $\cal{CP}$ problem to date, which has also led to numerous proposals for direct dark matter searches. Here we list but a fraction of the new (and old) ideas \cite{Budker:2013hfa,Graham:2013gfa,Rybka:2014cya,Sikivie:2013laa,Beck,Stadnik:2013raa,Sikivie:2014lha,McAllister:2015zcz,Hill:2015kva,Hill:2015vma,Kahn:2016aff,Barbieri:2016vwg, Arvanitaki:2014dfa} related to the axion search experiments.

On the other hand, one may also discuss a similar theta term in QED. It is normally assumed that a similar   $\theta_{\rm QED}$ parameter in the abelian Maxwell Electrodynamics is unphysical (if magnetic monopoles are absent), and can be always  
removed from the system. The arguments are precisely the same as given above and based on the observation that the $\theta_{\rm QED}$ term does not change the equation of motion. However, in contrast with QCD, the  topological mapping  for the abelian gauge group $\pi_3[U(1)]=0$ is trivial. This justifies the widely accepted understanding that $\theta_{\rm QED}$ does not affect any physical observables and can be safely removed from the theory.

While these arguments are indeed correct when the  theory is defined in an infinitely large 3+1 dimensional Minkowski space-time, 
it has been known for quite sometime that 
the $\theta$ parameter is in fact a physical parameter of the system when the theory is formulated  on a non-simply connected, compact manifold with 
non-trivial    $\pi_1[U(1)]=\mathbb{Z}$, see the original references  \cite{Witten:1995gf,Verlinde:1995mz} and review \cite{Olive:2000yy}. 

The  goal of the present work, however,  is not   an analysis of  the most  generic features of the Maxwell system such as the  duality relations in the presence of the $\theta$ parameter. Instead, mostly motivated by the dark matter axion search experiments,
here we only study simplest possible systems 
when the $\theta$ becomes a physically observable  parameter and discuss potential phenomenological consequences in idealized experiments. 
 Essentially, we want to search for   a system which would be highly sensitive to a  non-vanishing $\theta$.

   To achieve our  goal we consider a simplest possible  manifold such as ring with a single non-trivial    $\pi_1[U(1)]=\mathbb{Z}$.  We explicitly show why and how 
   the $\theta$ dependence emerges in such systems. What is more important, we explicitly compute all relevant factors related to this $\theta$ dependence. Roughly speaking, 
  the physics related to pure gauge configurations describing the topological sectors does not reduce to triviality when one removes all unphysical degrees of freedom as a result of gauge fixing in the course of the quantization of the Maxwell theory.
  The phenomena, in all respects, are very similar to the Aharonov-Bohm effect when the system is highly sensitive to pure gauge (but topologically nontrivial) configurations. 
  This is precisely a deep reason why   $\theta$ parameter\footnote{Here and in what follows we use $\theta$ rather than $\theta_{\rm QED}$ to simplify notations. It should not confuse the readers   as the only  $\theta$ parameter we have in the present work is the $\theta$ which couples exclusively to the gauge $E\&M$ fields, because we do not include effects related to gluons and fermions in our discussion.}  enters the physical observables in the axion Maxwell electrodynamics in full agreement with very generic arguments  \cite{Witten:1995gf,Verlinde:1995mz,Olive:2000yy}. Precisely these contributions   lead to the explicit $\theta$-dependent 
  effects, which cannot be formulated in terms of conventional propagating degrees of freedom. In fact, most of the  effects\footnote{The  exception is section \ref{winding} where a nontrivial topology is enforced by the external magnetic field rather than by a nontrivial non-simply connected manifold.}  which are subject of the present work are non-perturbative in nature as they enter the observables as $\exp(-1/e^2)$ and cannot be seen in perturbation theory. 
 Our explicit computations in next sections clarify  this claim. One should emphasize here that while parametrically $\exp(-1/e^2)$
 is exponentially small, numerically this factor could be of  order of one due to  a  special design of  a geometrical configuration.

Our result also implies that some physical observables, when considering a setup with nontrivial topology, can be proportional to $\theta$, as opposed to $\dot{\theta}$ as commonly assumed or discussed for perturbative computations. Precisely this feature has the  important applications in the axion search experiments where some observables are proportional to the  static time-independent  $\theta $, and, in general,  do not vanish even when $\dot{\theta}=0$.  
  
  Another important implication of our findings is that some physical observables may not be expressible in terms of propagating degrees of freedom, such as transverse photons. In other words, there are so-called ``non-dispersive" contributions to some correlation functions which are physical and observable, but cannot be expressed in terms of any   ``absorptive" contributions\footnote{\label{folk_theorem}It would contradict a ``folk theorem" that the $S$ matrix contains all information about all physical observables. We thank Mark Wise for providing and explaining this observation.} which carry  (through the dispersion relations)   only  the information about the   ``dispersive" portion of the correlation functions.

  The practical implication of this claim is that  there are some $\theta$-dependent contribution to the vacuum energy, which cannot be expressed in terms of any propagating degrees of freedom.  Precisely this type of non-perturbative contribution is related to the topological sectors of the axion Maxwell electrodynamics and the tunnelling transitions between them.
   The very same physical effects lead to the extra term in the vacuum pressure which was dubbed in  \cite{Cao:2013na}  as the   Topological Casimir Effect (TCE), representing  an additional non-perturbative contribution to the conventional Casimir Effect \cite{Casimir} and which cannot be expressed in terms of the physical propagating transverse photons. 
  
  The main goal of the present work is to elaborate
  on possible new $\theta$-dependent phenomena (mostly related to the axion search experiments) which 
  originated from the topological sectors in the Maxwell electrodynamics.
  
  The organization of this paper is as follows. In section \ref{sect:axion} we introduce some notations related to the axion physics.  We also   review a  variety of topological phenomena related to the $\theta$ term. In section \ref{Z_axion} we generalize 
  the construction of   \cite{Cao:2013na,Cao:2015uza}  of the topological portion of the partition function ${\cal{Z}}_{\rm top}(\tau, \theta ) $ to include the $\theta$ term in the Euclidean path integral formulation. As we already mentioned, the main goal of the present work    is not an analysis of some generic features (such as duality) of   
  the partition function ${\cal{Z}}_{\rm top}(\tau, \theta ) $ which is well known from previous studies \cite{Witten:1995gf,Verlinde:1995mz,Olive:2000yy}. Rather, we want to study a specific implementation on a very simple  geometry with specific and concrete parameters as an example, which would allow us to minimize the unavoidable    suppression factor  $\exp(-1/e^2)$   inevitably  occurring  as a result of the tunnelling transitions. 
  
 For this simple geometry  we explicitly compute  a number of observables in the Maxwell electrodynamics which depend on  $\theta$ rather than   $\dot{\theta}$, in agreement with known generic arguments \cite{Witten:1995gf,Verlinde:1995mz,Olive:2000yy} that the partition function ${\cal{Z}}_{\rm top}(\tau, \theta )$  itself depends on $\theta$. As the corresponding formulae play the  key role in our studies, we reproduce in section \ref{Hamiltonian} the relevant expressions  using the Hamiltonian approach formulated in the Minkowski space-time, in contrast with Euclidean formulation of the path integral. 
   Finally, in sections \ref{numerics}, \ref{winding}  we  propose a number of 
  experimental setups for possible axion search experiments. The crucial element in the design of the suggested  apparatus  
  is the presence of a topologically nontrivial configuration (such as a ring)  when the topologically  distinct sectors of the Maxwell system may manifest themselves and play the key role.  We conclude in Section \ref{conclusion} where we summarize the main results of our findings.

 \section{Axion $\theta$ field and variety of topological phenomena }\label{sect:axion}
We introduce, in the conventional way, the axion field $a(t) =f_a \theta (t)$ from which Maxwell equations receive corrections. 
The relevant term enters the Maxwell equation  as follows \cite{Sikivie:1983ip}
\be
\label{Maxwell}
\vec{\nabla}\times \vec{B}=\vec{j}+\frac{\vec{\partial E}}{\partial t}-g_{a\gamma\gamma}\vec{B}\frac{\partial a}{\partial t}, ~~~ g_{a\gamma\gamma}=\frac{K_{a\gamma\gamma} \alpha}{2\pi f_a}, 
\ee
where the spatial variation for the axion field  $\sim \vec{\nabla}a$ is consistently neglected because we assume it to be small and thus irrelevant for the present work. 
 The model dependent numerical coefficients are: $K_{a\gamma\gamma}\simeq 0.75$ for Dine-Fischler-Srednicki-Zhitnitsky (DFSZ) model and $K_{a\gamma\gamma}\simeq -1.92$ for Kim-Shifman-Vainshtein-Zakharov (KSVZ) model. 
 We treat the axion field as a classical coherent field \cite{Sikivie:2009fv} with normalization
 \be
 \label{axion}
 a(t)=\frac{\sqrt{2\rho_{\rm DM}}}{m_a}\cos (\omega_a t+\phi), ~~~~ \rho_{\rm DM}\simeq \frac{\rm 0.3~ GeV}{\rm cm^3}.
  \ee
 The coherence of the field is determined by the de Broglie wave length $ \lambda_D$ of the axions with a typical mass $m_a\sim 10^{-4} {\rm eV}$, 
 \be
 \label{lambda}
 \lambda_D=\frac{\hbar}{m_a v_a}\simeq 10\cdot \left(\frac{10^{-4} {\rm eV}}{m_a}\right) m .
 \ee
Our normalization corresponds to the assumption  that the axions  represent the dark matter  (or at least some portion of it) of the Universe. 

 The lower limit on the axion mass, as it is well known, is  
   determined by the requirement that  the axion contribution to the dark matter density does not  exceed the observed value $\Omega_{\rm dark}\approx 0.23$. There is a number of uncertainties in the corresponding estimates. The corresponding uncertainties are mostly due to the  remaining discrepancies between different groups on the computations  of the  axion production  rates due to the  different mechanisms such as  misalignment mechanism \cite{misalignment} versus  domain wall/string decays \cite{Hiramatsu:2012gg,Kawasaki:2014sqa,Fleury:2015aca}.   We shall not comment on these  subtleties by referring to the   original and review papers \cite{vanBibber:2006rb,Asztalos:2006kz,Raffelt:2006cw,Sikivie:2009fv,Rosenberg:2015kxa,Graham:2015ouw}.  If one takes for granted  that  the misalignment mechanism is dominant, which is normally assumed to be always  the case if  inflation occurs after the  Peccei-Quinn (PQ) phase transition, then the estimate is:
 \be
 \label{dm_axion}
 \Omega_{\rm (DM ~axion)}\simeq \left(\frac{6\cdot 10^{-6} {\rm eV}}{m_a}\right)^{\frac{7}{6}}, 
 \ee 
 for misalignment mechanism.
 This formula essentially states that the axion of mass $m_a\simeq 2\cdot 10^{-5}$ eV saturates the dark matter density observed today, while the axion mass in the range of $m_a\geq  10^{-4}$ eV contributes very little to the dark matter density. This claim, of course, is entirely based on estimate (\ref{dm_axion}) which accounts only for the axions directly produced by the misalignment mechanism.
 
There is another mechanism of the axion production when the PQ symmetry is broken after inflation as a result of the domain wall/string decays \cite{Hiramatsu:2012gg,Kawasaki:2014sqa,Fleury:2015aca}.  
 In this case the string-domain wall network produces a large number of axions such that   the dark matter density  might be saturated by   heavier   axion mass $m_a \simeq   10^{-4}$ eV, though there are large uncertainties and the  remaining discrepancies in the corresponding computations as we already
  mentioned\footnote{The only original comment we would like to make is as follows. It is normally assumed that
 the domain walls are the topological configurations which interpolate between physically distinct vacuum states. Generic  domain walls obviously cannot be produced if inflation occurs after the PQ phase transition, when there is  a unique physical  vacuum state within a horizon. The key point advocated in \cite{Liang:2016tqc} is that the very special $N_{\rm DW}=1$ domain walls  corresponding  to the field configurations   interpolating  between  {\it  topologically distinct} but physically {\it identical} states can be still produced even if the inflation occur after the PQ phase transition.  This is because the inflation cannot remove  these states  outside of the horizon. Somehow this option had been overlooked in previous studies on the subject.}.
 
 The range  $10^{-6} {\rm eV}\lesssim  m_a\lesssim   10^{-2} {\rm eV}$  is traditionally been regarded as the open mass window for the QCD axions, see recent review \cite{Graham:2015ouw}. In our  numerical estimates in the present work, such as (\ref{lambda}),   we use the central point $m_a\simeq 10^{-4}$ eV for illustrative purposes only.   This  choice is not dictated by any specific constraints related to an experimental design.  
 This ``insensitivity"    in our  
   current proposal  from  $m_a$ should be contrasted with conventional cavity-type experiments which  need to be fine-tuned for  a specific axion
   mass $m_a$.  The only requirement for our proposal is that the dark matter axion can be treated as  a coherent state (\ref{axion}) with a coherent length (\ref{lambda}) that is compatible with a typical size of the experimental configuration.

We now return to analysis of the Maxwell equations in the presence of the axion field given by eq.(\ref{Maxwell}).  
The physical meaning of extra term $\sim \dot{a}$ in (\ref{Maxwell})  is quite obvious: the axion generates the extra current $\vec{j}_a$ 
in the presence of the axion field
 \be
 \label{current}
 \vec{j}_a=  - \dot{\theta}~\frac{K_{a\gamma\gamma} \alpha}{2\pi} ~\vec{B}, ~~ \alpha\equiv \frac{e^2}{4\pi}.
 \ee
This anomalous current points  along magnetic field in contrast with ordinary $E\&M$, where the current is always orthogonal to $\vec{B}$.
Most of the recent proposals \cite{Budker:2013hfa,Graham:2013gfa,Rybka:2014cya,Sikivie:2013laa,Beck,Stadnik:2013raa,Sikivie:2014lha,McAllister:2015zcz,Hill:2015kva,Hill:2015vma,Barbieri:2016vwg, Arvanitaki:2014dfa,Kahn:2016aff} to detect the dark matter   axions    are precisely based on this extra current (\ref{current}).

  We would like to make a few comments on the unusual features of this current.
  First of all, the generation of the very same non-dissipating current (\ref{current}) in the presence of $\theta$
  has been very active area of research  in recent years.  However, it is with drastically different scale of order $\Lambda_{\rm QCD}$ instead of $m_a$. The main driving force for this activity stems from the ongoing experimental results 
  at RHIC (relativistic heavy ion collider) and the LHC (Large Hadron Collider), which can be interpreted as the observation of such anomalous current (\ref{current}). 
  
  The basic idea for such an interpretation can explained as follows. It has been suggested by \cite{Kharzeev:1998kz,Buckley:1999mv} that the so-called  $\theta_{\rm ind}$-domain can be formed in heavy ion collisions as a result of some non-equilibrium dynamics. This induced $\theta_{\rm ind}$ plays the same role as fundamental $\theta$ in (\ref{current}), 
 and leads to a number of $\cal{P}$ and $\cal{CP}$ odd effects, such as chiral magnetic effect, chiral vortical effect, and charge separation effect, to name just a few.   
 This field  of research initiated in \cite{Kharzeev:2007tn}  became a hot topic  in recent years as a result of many interesting theoretical and experimental advances,    see recent review papers \cite{Kharzeev:2009fn,Kharzeev:2015znc} on the subject. 
 
 For our present studies it is important to realize that  the anomalous non-dissipating current (\ref{current}) can be interpreted as a manifestation of the Witten's effect \cite{Witten:1979ey} when the magnetic monopole becomes an electrically  charged object, the dyon.  Indeed, as argued  in \cite{Kharzeev:2007tn} the electric field will always be induced in the presence of $\theta\neq 0$ if an external magnetic field is also present in the system, see (\ref{E}). This induced electric field $\la \vec{E}\ra_{\rm ind} \sim \theta \vec{B} $ will be generating the  non-dissipating current (\ref{current}) in plasma and may lead to a separation of charges\footnote{This separation of charges  observed  at RHIC and the LHC can be interpreted as the  manifestation of the charge separation effect \cite{Kharzeev:2007tn,Kharzeev:2009fn,Kharzeev:2015znc}. This effect can be also interpreted as the chiral magnetic effect (CME) represented by eq. (\ref{current}). The effect was dubbed as CME because $\dot{\theta}=\mu_5$ can be interpreted as the chiral chemical potential $\mu_5$. Relation $\dot{\theta}=\mu_5$ can be easily derived using $U(1)_A$ chiral time-dependent transformation in the path integral.}.

   A similar argument also suggests that a magnetic dipole moment will always generate  the electric dipole moment in the presence of the $\theta$ background. We elaborate on this argument in   section \ref{interpretation}, see eqs. (\ref{el_moment}), (\ref{m_e}), (\ref{m_e_witten}). Independent  explicit computations \cite{Hill:2015kva,Hill:2015vma} also support this argument. 
 
 Our final comment (on this large number of   topological phenomena emerging as a result of    the $\theta$ term) is as follows. It is commonly  assumed that the physics related to the $\theta$ parameter in electrodynamics must decouple in the limit $\dot{\theta}\rightarrow 0$, so that all physical phenomena must be $\theta$ independent if $\dot{\theta}= 0$. The conventional argument is often based on observation that the $\theta$ term in this limit can be represented as the total derivative in the action,  and, therefore, cannot change the equation of motion or influence physical observations, see \cite{Kharzeev:2009fn} for review.  As we already mentioned in the Introduction, it has been known for sometime   \cite{Witten:1995gf,Verlinde:1995mz,Olive:2000yy} why these arguments do not hold for  a system  which is   formulated on a nontrivial manifold\footnote{\label{Gribov}This phenomena is in fact related to the so-called Gribov ambiguities which are well known to  emerge in non-abelian gauge theories.  These ambiguities  also appear in  the Maxwell electrodynamics when it is quantized on a non-simply connected compact manifold,   see some comments and relevant references on the subject in   \cite{Cao:2013na}. }.
Our   computations below which are performed  on a  simple, but topologically nontrivial  manifold $S^1$  explicitly show how and why this $\theta$ dependence emerges.

\section{Topological partition function in presence of the axion.}\label{Z_axion}
We want to construct the partition function  ${\cal{Z}}_{\rm top} (\theta)$ in the presence of the axion field $\theta (t)$. Mathematically, we consider the usual 3-dimensional space with an infinitely long cylindrical portion removed, the space of consideration is thus homotopic to $S^1$. For simplicity, we align the axis of the (rectangular) cylinder with the z-direction. The single periodic direction is provided by the $S^1$ factor of the cylinder. Certainly, the real spatial topology for Minkowski space is trivial, and one may argue that such constructions can never be physically realized unless the non-trivial topology is provided by the actual Universe. Such will be the case if the Universe is a torus or has handles from wormholes. We here make a few arguments that although change in real spatial topology cannot be controlled in an experimental setting, it can be effectively realized in real space at low energies. For instance, while a conducting ring does not change the actual spatial topology, it does enforce the $S^1$ topology for some low energy electrons below some binding energy. 
The accumulation of the Aharonov Bohm phase with  non-contractible path along $S^1$ is another example when the topology can be enforced by physical consideration, see additional  comments below. 

For our present work, one possible realization is by inserting an actual cylinder with cross-sectional size a lot smaller than its length along $z$. We claim that the periodic boundary condition is indeed enforced up to some energy cut off provided by the material, as in the case of Casimir effects \cite{Bordag:2009zzd}. As expected, the imposed material boundary becomes transparent for modes above such cut offs. Nevertheless, the boundary condition is enforced in the regime in which vacuum Casimir effect is dominant. The validity of such arguments are indeed supported by the actual measurement of Casimir pressure in vacuum. 

As expected, the correction to Casimir pressure from such non-perturbative effects  is a few orders of magnitude smaller than the current detection limit \cite{Cao:2013na}. Nevertheless, by designing an intrument with many identical units, some related effects such as induced dipole moments can be magnified. We also discuss the effects when a strong external field is threading through a region of space. In such cases, non-trivial topology can also enter even in the absence of material boundaries. By threading a sufficiently strong external field in some region, it is sufficient to consider the zero field region for low energy effects. Such region will be homotopic to $S^1$ as desired. Processes that correspond to unwinding have to pass through the non-zero flux region which has a higher energy proportional to the sqaure of the strengths of the applied fields. However, we will not give explicit computations in this paper to demonstrate the explicity equivalence in the low energy limit, which involves non-perturbative corrections and is highly non-trivial as a topic by itself. In practice, of course, the cylinder is never infinitely long and external fluxes will not extend to infinity. However, topology is enforced for sufficiently long cylinders or extended fluxes, when we consider thin spatial slices cutting through the cylinder which are effectively 2-dimensional and are far from the ends. A well-known example will be the familiar Aharonov-Bohm setup where interference pattern can be detected despite the finite size of the solenoid. This sets the range of validity for the computations throughout this paper.

Working under the assumptions that such non-trivial topology can indeed be effectively realized, we now follow \cite{Cao:2013na,Cao:2015uza} in constructing the topological partition function, which is sensitive to the topological sectors of the system.  As a crucial observation from previous work, we note that the gauge field $A^{\mu}$ is only periodic up to a large gauge transformation in the periodic direction, and results in formation of the topological sectors $|k\ra$. In this section we use conventional Euclidean path integral formulation and then reproduce the key results using the Hamiltonian analysis in section \ref{Hamiltonian}. 

For the sake of clarity, we briefly review the fluctuations  due to the magnetic fluxes \cite{Cao:2013na} in section \ref{construction} and neglect the possible electric fluxes discussed in ref. \cite{Cao:2015uza} in the periodic direction. 
In subsection \ref{Z_theta} we generalize this construction to include the axion $\theta$ field.
Finally, in subsection \ref{interpretation} we argue that the obtained results can be interpreted as the manifestation of the Wittens' effect  \cite{Witten:1979ey} when the magnetic monopole becomes electrically  charged object, the dyon. 

This construction explicitly shows that
the $\theta$ is a physical parameter of the system even though it enters the Lagrangian (\ref{theta}) with operator which can be represented as a total derivative. One can explicitly see that all effects which are subject of the present work are non-perturbative in nature as they enter the partition function (\ref{Z_eff}) as $\exp(-1/e^2)$ and cannot be seen in perturbation theory.

  \subsection{Maxwell system on a compact manifold}\label{construction}

 In what follows we simplify our analysis by considering   a simplest  case with contributions from winding topological sectors $|k\ra$    in the $z$-direction only.  The classical instanton configuration in Euclidean space  which describes the corresponding tunnelling transitions can be represented as follows:
\be
\label{toppot4d}
A^{\mu}_{\rm top} = \left(0 ,~ -\frac{\pi k}{e L_{1} L_{2}} x_2 ,~ \frac{\pi k}{e L_{1} L_{2}} x_1 ,~ 0 \right),
\ee  
where $k$ is the winding number that labels the topological sector, and $L_{1}$, $L_{2}$ are the dimensions of the rectangular cylinder in the x and y-directions respectively. The height of the cylinder is $L_3$. This terminology (``instanton") is adapted   from  similar 
 studies in 2d QED    \cite{Cao:2013na} where corresponding configuration in $A_0=0$ gauge describe the interpolation between pure gauge vacuum winding states $|k\ra$.  
 
 The physical meaning of the configurations (\ref{toppot4d}) is that they describe the tunnelling processes which occur in the system between different winding   sectors $|k\ra$. For relatively small systems and finite temperature $\beta$ the probability for the tunnelling processes is not small, and must be taken into account to describe  the physical vacuum states.

A key observation from \cite{Cao:2013na} is that the  topological portion ${\cal{Z}}_{\rm top}$   decouples from quantum fluctuations,  ${\cal{Z}} = {\cal{Z}}_{\rm quant} \times {\cal{Z}}_{\rm top}$ such that the quantum fluctuations from propagating photons do not depend on topological sector $k$ and can be computed in topologically trivial sector $k=0$.
Indeed,  the cross term 
\be
\int \dd^4 x~ \vec{B} \cdot \vec{B}_{\rm top} = \frac{2 \pi k}{e L_{1} L_{2}} \int \dd^4 x~ B_{z} = 0 
\label{decouple}
\ee
vanishes  because the magnetic portion of quantum fluctuations in the $z$-direction, represented by $B_{z} = \partial_{x} A_{y}  - \partial_{y} A_{x} $, is a periodic function as   $\vec{A} $ is periodic over the domain of integration. This technical remark in fact greatly simplifies our  analysis.

The classical action for configuration in the presence of the uniform static external magnetic field $B_{z}^{\rm ext}$ therefore takes the form 
\be
\label{B_ext}
\frac{1}{2}\int \dd^4 x  \left(\vec{B}_{\rm ext} + \vec{B}_{\rm top}\right)^2=  \pi^2\tau\left(k+\frac{\theta_{\rm eff}}{2\pi} \right)^2
\ee
where parameter $\tau$ is defined as  $\tau = 2\beta L_3/e^2L_1L_2$, while    the effective theta parameter $\theta_{\rm eff} \equiv e L_1L_2 B^z_{\rm ext}$ is expressed in terms of the external magnetic field $B^z_{\rm ext}$.
Therefore, the partition function in the presence of the uniform magnetic field  is given by  \cite{Cao:2013na,Zhitnitsky:2015fpa} 
\be 
\label{Z_eff}
  {\cal{Z}}_{\rm top}(\tau, \theta_{\rm eff})
 =\sqrt{\pi\tau} \sum_{k \in \mathbb{Z}} \exp\left[-\pi^2\tau \left(k+\frac{\theta_{\rm eff}}{2\pi}\right)^2\right].~~
\ee
This system in what follows will be referred as the topological vacuum ($\cal{TV}$) because the  propagating degrees of freedom, the photons with two transverse polarizations,   
completely decouple from  ${\cal{Z}}_{\rm top}(\tau, \theta_{\rm eff})$.

  The dual representation for the partition function is obtained by applying the Poisson summation formula  such that (\ref{Z_eff}) becomes 
  \be 
\label{Z_dual1}
  {\cal{Z}}_{\rm top}(\tau, \theta_{\rm eff})
  = \sum_{n\in \mathbb{Z}} \exp\left[-\frac{n^2}{\tau}+in\cdot\theta_{\rm eff}\right]. 
  \ee
 Formula (\ref{Z_dual1})  justifies our notation for  the effective theta parameter $\theta_{\rm eff}$ as it enters the partition function in combination with integer number $n$. One should emphasize that integer  number $n$ in the dual representation (\ref{Z_dual1}) is not the integer magnetic flux $k$.    Furthermore,  the $\theta_{\rm eff}$ parameter which enters (\ref{Z_eff}, \ref{Z_dual1}) is not a fundamental $\theta$ parameter which is normally introduced into the Lagrangian  in front of  $\vec{E}\cdot\vec{B}$ operator. This fundamental $\theta$ term  describing the physical axion field will be 
 introduced in next section. Rather, this parameter  $\theta_{\rm eff}$ should be understood as an effective parameter representing the construction of the  $|\theta_{\rm eff}\ra$ state for each 2-dimensional slice with non-trivial $\pi_1[U(1)]$ in the four dimensional system.

 \subsection{Euclidean partition function in the presence of $\theta$}\label{Z_theta}
 Our goal now is to generalize formula (\ref{Z_eff}) to include the fundamental $\theta (t)$ into the partition function ${\cal{Z}}_{\rm top}(\tau, \theta, \theta_{\rm eff})$. As such, we insert an extra $\theta$ term into  the Euclidean action (\ref{B_ext}),

  \be
\label{theta}
S_E=\int \dd^4 x  \left[\frac{1}{2}\vec{B}^2+\frac{1}{2}\vec{E}^2+ i \frac{K_{a\gamma\gamma} \alpha}{\pi} {\theta} \vec{E}\cdot\vec{B}\right]. ~~~
 \ee
Note that our normalization for the topological term is different from conventional definition of $\theta$ in the so-called ``axion electrodynamics" in condensed matter physics  which corresponds to $K_{a\gamma\gamma}=1$ in our equation (\ref{theta}).  Another comments is related to complex factor $i$ in the definition (\ref{theta}). This is the result of the Euclidean signature when the  time and electric field are  in fact imaginary variables. 

Variation of the action $S_E$ with respect to electric field $\delta/\delta\vec{E}$ returns 
\be
\label{E}
\la \vec{E}\ra_{\rm ind}=-i  \frac{K_{a\gamma\gamma} \alpha}{\pi} {\theta} \left( \vec{B}_{\rm ext} + \la \vec{B}\ra_{\rm ind}\right),
\ee
 which implies that the electric field will be always induced in the presence of $\theta$ term.
 In what follows we want to simplify all formulae and consider the limit of $\theta\rightarrow 0$, in which case
 the computations of the rhs $  \la \vec{B}\ra_{\rm ind}$ entering (\ref{E}) can be carried out 
 with the partition function (\ref{Z_eff}) computed at $\theta=0$. In this approximation the induced electric field assumes the form
  \begin{align}
  \label{E_ind}
  &\la {E^z}\ra_{\rm ind}=-i  {\theta}\cdot  \frac{K_{a\gamma\gamma} \alpha}{\pi}\cdot  \frac{\sqrt{\tau \pi}}{{\cal Z}_{\rm top}(\tau, \theta_{\rm eff})}\\
  & \times \sum_{k \in \mathbb{Z}} (B^z_{\rm ext} +\frac{2 \pi k}{e L_1 L_2}) \exp[-\tau \pi^2(k+\frac{\theta_{\rm eff}}{2\pi})^2].\nonumber
  \end{align}
  This formula (written in Euclidean metric) represents the main result of this section.  It  shows that the   observables explicitly depend
  on $\theta$ through  a parametrically small suppression factor $\sim \exp(-1/e^2)$, in agreement with generic arguments presented in the Introduction.  In next section we interpret the obtained result 
   from a different perspective. 
 
On the other hand, we can also consider the dual picture whereby an external electric field $E_{\rm ext}$ is applied and a magnetic flux is induced from nonzero $\theta$. Although not the physical configuration we consider in this paper, we note the duality becomes manifest, when the system is endowed with an additional $z$-periodicity. As such, one similarly obtains, in Euclidean metric

\be
\label{dual_indfield}
\langle B^z\rangle_{\rm ind} = -i\theta\frac{K_{a\gamma\gamma}\alpha}{\pi} \left(E_{\rm ext}^z +\la E^z\ra_{\rm ind}\right),
\ee
whereby a magnetic flux is induced by external electric field, as expected. The induced electric field $\la E^z\ra_{\rm ind}$ in this expression is related to the electric fluxes, which can be in principle computed \cite{Cao:2015uza}, is dropped here in the current configuration where $z$-periodicity is lacking.

A few comments regarding (\ref{E}) are in order. Naively, (\ref{E}) seems to suggest that the electric  field $\la \vec{E}\ra_{\rm ind}$ will be always induced regardless of the topological features of the space, which we claim plays the crucial role. Nevertheless, this  naive objection   is incorrect as we argue below.
Just as the electric charge  will be always induced in the background of the   magnetic monopole  (see eq. (\ref{m_e_witten}) and comments after this formula below), similar non-triviality is induced by a magnetic flux (string) which is always accompanied at large distances by a pure gauge    (but topologically nontrivial) vector potential, see section \ref{winding} and eq. (\ref{theta1}) with  details analysis. Precisely this pure gauge but topologically nontrivial vector potential is responsible for  a nontrivial mapping 
$\pi_1[U(1)]=\mathbb{Z}$ in configurational space, which plays the key role in the entire construction.  In other words, the external magnetic flux itself, like the case of Aharonov-Bohm effect, is providing (and enforcing) the required non-trivial topology. Therefore, for external field with finite extent, the $\theta$-term does not reduce to zero on the boundary as usual, but rather generates a number of  nontrivial effects. This $\theta$ dependence in (\ref{E})  emerges, of course,  because the object of our studies is not the QED vacuum (where $\theta$ parameter indeed can be safely removed), but rather a heavy sector with non-vanishing magnetic flux, see   section \ref{winding}  with  additional explanations and comments on this matter. 

 \subsection{Interpretations}\label{interpretation} 
 First of all, the expression for the induced electric field (\ref{E_ind}) is written in Euclidean metric where all path integral computations 
 related to the tunnelling transitions are normally performed. As usual, we assume that analytical continuation is valid and, therefore, the same formula holds  for the physical electric field in Minkowski space-time, which is obtained by  removing complex ``$i$'' in front of eq. (\ref{E_ind}).   
 
 Our next comment is about the interpretation of the obtained formula (\ref{E_ind}).  First, we would like to interpret the induced magnetic field $ \la {B}\ra_{\rm ind}$ in terms of the induced magnetic moment in each given topological sector $k$ 
 \begin{align}
  \label{mag_moment}
  & \la m_{\rm ind}\ra=-\left(B^z_{\rm ext} +\la B\ra_{\rm ind}\right) L_1 L_2 L_3 =- \frac{\sqrt{\tau \pi}}{{\cal Z}_{\rm top}} \\
  &\times \frac{2\pi L_3}{e} \sum_{k \in \mathbb{Z}} \left(\frac{\theta_{\rm eff}}{2\pi}+k\right) \exp[-\tau \pi^2(k+\frac{\theta_{\rm eff}}{2\pi})^2]. \nonumber
  \end{align}
 Furthermore,  the corresponding magnetic moment in  $k$ sector can be also understood in terms of the induced non-dissipating persistent currents which flow along infinitely thin boundary of the system  as  discussed in \cite{Zhitnitsky:2015fpa}. 

Novel element emerges as a result of   $\theta$ parameter. In this case the induced electric field (\ref{E_ind}) can be interpreted in terms of the induced dipole moment
 \begin{align}
  \label{el_moment}
  & \la d\ra_{\rm ind}=-\la E\ra_{\rm ind} L_1 L_2 L_3 ={\theta}\cdot  \frac{K_{a\gamma\gamma} \alpha}{\pi} \frac{\sqrt{\tau \pi}}{{\cal Z}_{\rm top}(\tau, \theta_{\rm eff})} \\
  &\times \frac{2\pi L_3}{e} \sum_{k \in \mathbb{Z}} \left(\frac{\theta_{\rm eff}}{2\pi}+k\right) \exp[-\tau \pi^2(k+\frac{\theta_{\rm eff}}{2\pi})^2], \nonumber
  \end{align}
where we adopted the Minkowski signature for the physical electric field in contrast with expression (\ref{E_ind}) derived in Euclidean space-time. The same expression for the electric dipole moment can be thought as accumulation of the charges on the metallic plates 
of the area  $L_1L_2$ and separated by distance $L_3$ along $z$. 

Comparison between (\ref{mag_moment}) and (\ref{el_moment})
suggests that the induced electric field in the presence of $\theta$ can be thought as the Witten's effect  as the electric and magnetic dipole moments are related:
\be
\label{m_e}
 \la d\ra_{\rm ind}= -{\theta}\cdot  \frac{K_{a\gamma\gamma} \alpha}{\pi}  \la m_{\rm ind}\ra
\ee
which    obviously resembles the Witten's  relation if one represents the magnetic moment as $ \la m_{\rm ind}\ra =g L_3$ where $g$ is the magnetic charge.  As the magnetic charge $g$ is quantized, $g=\frac{2\pi}{e}$,  formula (\ref{m_e}) can be rewritten as
 \be
\label{m_e_witten}
 \la d\ra_{\rm ind}= -{\theta}\cdot  \frac{K_{a\gamma\gamma}e^2}{4\pi^2}  \frac{2\pi L_3}{e}=-\left(\frac{e\theta }{2\pi}\right)  L_3\cdot  K_{a\gamma\gamma}.~~
\ee
 This formula can be obviously interpreted as the Witten's effect when the magnetic dipole $\la m_{\rm ind}\ra$ becomes also an electric dipole  $\la d\ra_{\rm ind}$ with the moment
 determined by the electric charges $e'= - ({e\theta }/{2\pi}) K_{a\gamma\gamma}$ which precisely coincides with the Witten's expression for $e'= - ({e\theta }/{2\pi})$ if one uses the conventional normalization  for the $\theta$ term with $K_{a\gamma\gamma}=1$ according to  \cite{Witten:1979ey}.
 
 The discussions above strongly suggest that the the effects in the bulk of the system  can be represented in terms of the boundary effects: 
 boundary induced currents, or boundary induced charges. It is not really an unexpected property  of the system as we previously argued that this topological vacuum ($\cal{TV}$) is, in many respects, similar to a topologically ordered system \cite{Zhitnitsky:2013hba,Zhitnitsky:2014dra}. Therefore, the representation of the effect (\ref{E_ind}) in terms of the   boundary sources is another manifestation of the   property which is normally attributed to a system which belongs to  a   topologically ordered phase. 
 
 Formula (\ref{el_moment}) plays the crucial role in our analysis  in section \ref{numerics} where we describe a possible experimental set-up for the axion search experiment. It is explicitly proportional to the axion field $\theta(t)$, instead of $\dot{\theta}(t)$. This formula 
  also  explicitly  demonstrates  that  $\theta_{\rm QED}=$ constant  is a fundamental and physically observable parameter of the theory when  the system is formulated on a nontrivial manifold, in full agreement with very generic arguments of refs.
  \cite{Witten:1995gf,Verlinde:1995mz,Olive:2000yy}.

 \section{Hamiltonian Approach}\label{Hamiltonian}

 In this section, we reproduce (\ref{el_moment}) using the Hamiltonian approach and confirm the results from the path-integral calculation.  To be more self-contained, we briefly review some of the derivation in the Hamiltonian approach.

Following \cite{Chen-Lee:2010}, we can write down the zero-mode contribution to partition function on a 3-torus. Note that this is different and somewhat more complicated than the cylindrical system we constructed previously. We will find that we recover the cylindrical result by limiting to a trivial winding sector. We will find that the system which has a slightly more complicated topology contain other physical effects absent in the cylindrical case. As argued previously in  \cite{Zhitnitsky:2014dra}, the full partition function will contain six different fluxes where an integer electric and magnetic flux will be threading each independent direction. For the sake of simplicity, we focus on fluxes along $z$, although it is straightforward to extend to the full 3-torus solution where it is a triple copy of the partial partition function we consider here. 

Knowing the full system Hamiltonian $H$ given by Maxwell theory, the full partition function is given by 

\begin{equation}
\cal{Z}=\Tr[ \exp(-\beta H)]\sim \cal{Z}_{\rm top}\times \cal{Z}_{\rm quant},
\end{equation}
where $Z_{\rm top}=\Tr[\exp(-\beta H_{\rm top})]$ and $H_{\rm top}$ is the zero mode component of the total Hamiltonian.

Note that we can separate the partition function into its zero mode contribution $H_{\rm top}$ and the higher fluctuations such as physical photons, captured in $\cal{Z}_{\rm quant}$. This decoupling has been previously demonstrated  \cite{Cao:2013na}. Because here we are only interested in the topological part associated with the zero mode, it suffices to drop the conventional photon partition function $\cal{Z}_{\rm quant}$.

To start, we recall that for the system for a non-trivial topology admits different winding states that are related by a large gauge transformation. The true theta vacuum is labelled by a free parameter $\tilde{\theta}$, whereby the one has to take a superposition of winding states,

\begin{equation}
|\tilde{\theta}\rangle = \sum_{\tilde{n}=-\infty}^{\infty}\exp(i\tilde{\theta} \tilde{n})|\tilde{n}\rangle
\end{equation}
where each winding state is labelled by integer $\tilde{n}$. 

It was shown in  \cite{Cao:2013na,Cao:2015uza} that integer-valued electric and magnetic fluxes will thread the z-direction for the $\tilde{\theta}=0$ case. Our first step here is consider the $\tilde{\theta}$-dependence. Here $\tilde{\theta}=-\theta\alpha K_{a\gamma\gamma}/\pi$ is from the axion coupling term.

The topological contribution is given by partition function
\be
\cal{Z}_{\rm top}&\sim& \int^{2\pi}_{0}d\phi\int_{-\infty}^{\infty}\sum_{\tilde{m}}\sum_{\tilde{n}}\langle \frac{\phi}{2\pi}+\tilde{m}|\exp(-\beta H_{\rm top})|\frac{\phi}{2\pi}+\tilde{n}\rangle \nonumber \\
&\times& \exp(i(\tilde{m}-\tilde{n})\tilde{\theta}).
\ee
Equivalently, we note that $\exp(i(\tilde{m}-\tilde{n})\tilde{\theta})=\exp(i\tilde{\theta} \int d^4x \mathcal{L}_{\rm CS})$ is captured by the theta term, whose exponent is nothing but the abelian Chern-Simons action
\begin{equation}
\int d^4x \mathcal{L}_{\rm CS}=\frac{1}{8\pi^2}\int \mathcal{F}\wedge\mathcal{F},
\end{equation}
where $\mathcal{F}=d\mathcal{A}$ is the usual 2-form associated with the electromagnetic field strength. 
This has been computed for a 3-torus in  \cite{Chen-Lee:2010} and is nothing but $8\pi^2mn/e^2$, where $m,n$ are integers.

As such, one can insert identities $\int dl |\ell\rangle\langle l|$ into the expression where $|l\rangle$ is an eigenstate of the conjugate momentum zero mode $\tilde{E}^z(0)L_1L_2/e$ with eigenvalue $l$. 
Evaluate and one obtains partition function  \cite{Chen-Lee:2010}
\be
\cal{Z}_{\rm top}(\tilde{\theta})&\sim& \int^{2\pi}_{0}d\phi\int_{-\infty}^{\infty}dl\sum_{m,n}\langle \frac{\phi}{2\pi}+n|l\rangle\langle l|\exp(-\beta H_{\rm top})|l\rangle  \nonumber \\
&\times &\langle l |\frac{\phi}{2\pi}\rangle\exp(i mn\tilde{\theta}) 
\sim\sum_{m,n} \int \frac{dl}{2\pi}\delta \left(\frac{m\tilde{\theta}}{2\pi}+l+n\right) \nonumber \\
&\times& \exp\left(-\frac{\beta V}{2}\left[(\frac{el}{L_1L_2})^2+(\frac{2\pi m}{eL_1L_2})^2\right]\right)
\ee
up to constant normalization. Evaluating the element $\langle l|\exp(-\beta H_{\rm top})|l\rangle$ yields $\exp(-\beta V/2 (el/L_1L_2)^2+\Phi_B^2)$ where the integer magnetic flux $\Phi_B^2\sim m^2$ was given in \cite{Cao:2013na}. As expected, the partition function is invariant under $\tilde{\theta}\rightarrow \tilde{\theta}+2\pi$ and nonzero theta introduces electromagnetic mixing.

Although the induced field can be trivially computed from the partition function, unsurprisingly, the $\theta$-dependent induced field vanishes for all values of $\tilde{\theta}\sim\theta_{\rm QED}$ due to cancellation in the summation.
As before, we need to apply an external field to break the invariance of the exponential weight factor under $m\rightarrow -m$ and the same thing for $n$.  

For instance, we can shift magnetic flux by adding an external $B$ field in the $z$-direction. This shifts the magnetic flux, such that $2\pi m\rightarrow 2\pi m+eL_1L_2 B_{\rm ext}$. Setting $\theta_{\rm eff}=eL_1L_2B_{\rm ext}$, we have for total flux density

\be
\label{E_Hamiltonian}
&&\langle E\rangle \sim \sum_{m,n} \left[(\frac{\theta_{\rm eff}}{2\pi}+m)\frac{\tilde{\theta}}{L_1L_2e}+\frac{2\pi n}{e\beta L_3}\right]\\
&&\times\exp\left[-\frac{1}{\eta}\left(n+(m+\frac{\theta_{\rm eff}}{2\pi})\frac{\tilde{\theta}}{2\pi}\right)^2-\tau\pi^2\left(m+\frac{\theta_{\rm eff}}{2\pi}\right)^2\right],\nonumber
\ee
where $\eta=2L_1L_2/e^2\beta L_3$ and $\tau = 2\beta L_3/e^2L_1L_2$ and  we omit constant normalization factors proportional to $1/Z_{\rm top}$. Magnetic moment can also be computed in a similar fashion. Therefore, for small $\theta$ perturbations, as is the case for coherent dark-matter axions, we induce time-dependent electric flux

\begin{equation}
 \langle E(t)\rangle =  -\frac{\alpha K_{a\gamma\gamma}}{\pi}\gamma (\theta_{\rm{eff}},\tau) \theta(t)+\mathcal{O}(\theta^2)
\end{equation}
where it is important to note $\theta_{\rm{eff}}\sim B_{\rm{ext}}$ is external magnetic flux we apply \textit{parallel} to the induced E-flux. 

Here 
\begin{align}
&\gamma(\theta_{\rm{eff}},\tau)\equiv\frac{\partial \langle E(\tilde{\theta})\rangle}{\partial \tilde{\theta}}\Big|_{\tilde{\theta}=0} \\
&= \frac{2\pi}{eL_1L_2} \langle \frac{\theta_{\rm{eff}}}{2\pi}+m\rangle\Big|_{\tilde{\theta}=0} -\frac{2}{e L_3\beta \eta }\langle  n^2 (m+\frac{\theta_{\rm{eff}}}{2\pi})\rangle\Big|_{\tilde{\theta}=0} . \nonumber
\end{align}
Expectation values $\langle \dots \rangle$ are understood to be with respect to the topological partition function.

Note that we recover the case where the spatial topology has only one periodic direction, $S^1\times I\times I$, by only considering the $n=0$ sector, because there is no $z$-periodicity in our configuration, which is needed for non-trivial electric flux. It precisely agrees with equations (\ref{E_ind}) and (\ref{el_moment}) derived in  section \ref{Z_axion} using drastically different technique. Formula (\ref{E_Hamiltonian}) is in fact a more general expression  than  eq. (\ref{E_ind}) derived previously because  in formula (\ref{E_Hamiltonian}) the axion field $\tilde{\theta}$ is not assumed to be small, and furthermore, electric fluxes studied in  \cite{Cao:2015uza} which describe  $n\neq 0$  sectors, are included into the formula  (\ref{E_Hamiltonian}). Reproducing of eq. (\ref{E_ind}) using a different approach also adds to our confidence that the derivation is indeed correct.

 \section{Experimental Setup and Numerical Estimates for $|\theta_{\rm eff}\ra$ states}\label{numerics}
We give a  few simple numerical estimates related to the induced field   (\ref{E_ind}) and induced dipole moment (\ref{el_moment}) as a result of the external $\theta(t)$ field which is identified with the dark matter axion  (\ref{axion}).   We also want to present proposals on possible experimental setups which would allow to measure the induced field $\la \vec{E}\ra_{\rm ind}$ resulting from the DM axion entering a proposed axion detector.  

For coherent DM axions which are believed to have the wavelength in the order of a few meters, we can still operate in the adiabatic regime for a device with size on the scale of mm. The typical parameters in this section are: the axion mass $m_a\simeq 10^{-4}$ eV which corresponds to the typical frequencies $\nu\sim 20 $~GHz.  A typical  external magnetic field to be considered in this section  should be the same order of magnitude as 
magnetic field in the quantum fluxes describing the tunnelling events. Numerically, the magnetic field is   quite small, 
$B_{\rm ext}\sim \theta_{\rm eff}/(eL_1L_2)\sim 4\cdot 10^{-5} G$ for a typical mm size  samples. We also require the temperature of the system is not too low, $T \geq   10K$ which corresponds to  $\beta \leq  0.2 {\rm mm}$. At the same time, the    temperature cannot be too high as  the Aharonov-Bohm coherence must be maintained in the entire system\footnote{\label{AB}One should remark here that there are related effects  when  the entire system  can   maintain   the  Aharonov-Bohm phase coherence  at    very  high temperature $T\simeq  79 $ K \cite{persistent-temp}. It is obviously more than sufficient   for our purposes.}. 
 These parameters guarantee that the key dimensionless 
parameter  $\tau \lesssim 1$ and the tunnelling transitions  will  occur without exponential suppression. These tunnelling transitions select a specific 
$|\theta_{\rm eff}\ra $- state. Our classification in the present section is based on this classification scheme.  

In next section \ref{winding} we consider another set of parameters when the requirement on small $\tau$ can be dropped. In this case the tunnelling transitions do not occur as they are strongly suppressed, and our classification scheme will be based on the winding number topological  sectors $|k\ra$ rather than on $|\theta_{\rm eff}\ra $- classification which represents the superposition of the winding $|k\ra$ states. Nevertheless, the equations (\ref{E})
and (\ref{dual_indfield}) still hold even when tunnelling transitions are suppressed. We further elaborate on this classification scheme in section   \ref{winding}.

\exclude{
is in the few GHz range with $\dot{\theta} \sim 10^{-6} eV \ll \Delta E \approx 10^{-3} \rm eV \sim 10^{-5} \rm eV$, where $\Delta E$ is the gap in the effective transition energy \cite{Yao:2016bps} in the mm range. 
}
For subsections \ref{subsec:photonemission} and \ref{subsec:dipolemeasurement}, we consider the simple configuration of applying the external flux using a single cylindrical solenoid with mm size. An external magnetic field is applied parallel to the principle axis by the solenoid and an electric field (parallel to the external magnetic field) will be  induced in the presence of nonzero $\theta$.  For subsequent sections \ref{subsec:potentialdiff} and \ref{subsec:transientcurrent}, we put additional plates near the ends of the cylinder and connect them by a (super)conducting wire.

Note that while each device, which we study in the following subsections, only yields a minute amount of observable effect, the small scale of the devices allows in principle an amplification by considering a large number, $N$, of such  devices. For instance, for a cubic meter size detector, which is still well below the axion wavelength such that coherence is maintained, one can in principle pack $N^3$ mm scale construction of each unit and attains a factor of $\sim 10^9$ amplification.

\subsection{ The Dipole Moment}\label{subsec:dipolemeasurement}
We start our numerical estimates with magnetic dipole moment given by (\ref{mag_moment}).
If somehow we    manage to   adjust parameters of the system such that $\tau\leq 1$    than the magnitude of  $ \langle {m}^z_{\rm ind} \rangle$  from eq. (\ref{mag_moment}) is determined by parameter $2\pi L_3/e$ such that
\be
\label{estimate}
  \langle {m}^z_{\rm ind} \rangle \sim \frac{2\pi L_3}{e} \sim 1.5\cdot 10^{11} \left(\frac{e \cdot {\rm cm}^2}{{\rm s}}\right)\cdot\left(\frac{L_3}{1 {\rm mm}}\right).
\ee
By expressing  the magnetic moment$  \langle {m}^z_{\rm ind} \rangle $ in units  (\ref{estimate}) we want to emphasize 
that this   magnetic moment can be interpreted as the generation of the persistent current along the ring as described in 
\cite{Zhitnitsky:2015fpa}. In other words we interpret  $\langle {m}^z_{\rm ind} \rangle$  as a macroscopically large magnetic moment which is generated by coherent non-dissipating surface current $\la J\ra$ flowing along the ring and measured in units $e/s$. In this case one can represent  $ \langle {m}^z_{\rm ind} \rangle = L_1L_2 \la J\ra$ which explains our representation in form (\ref{estimate}).

The value (\ref{estimate}) should be compared numerically with Bohr magneton for a single electron represented in the same units, 
\be
\mu_B=\frac{e\hbar}{2 m_e}\simeq 0.6  \left(\frac{e \cdot {\rm cm}^2}{{\rm s}}\right), ~~ \frac{ \langle {m}^z_{\rm ind} \rangle}{\mu_B}\sim 10^{11} .
\ee
The comparison between the two numbers can be interpreted that our system effectively describes $\sim 10^{11}$ degrees of freedom which coherently produce a macroscopically large magnetic moment (\ref{estimate}) and coherent persistent current $\la J\ra$.
 This enhancement is accompanied by another  large factor $N^3\sim 10^9$ mentioned above. These two large factors represent  the maximum enhancement which can be achieved for a coherent axion field with $\lambda_D\sim 1$m.  The enhancement factor $10^{11+9}$ is the same order of magnitude which is normally discussed in other  axion search experiments \cite{Budker:2013hfa,Graham:2013gfa,Rybka:2014cya,Sikivie:2013laa,Beck,Stadnik:2013raa,Sikivie:2014lha,McAllister:2015zcz,Hill:2015kva,Hill:2015vma,Barbieri:2016vwg, Arvanitaki:2014dfa,Kahn:2016aff} based on coupling of the axion with the matter fields. 

  The crucial  element in our  estimate (\ref{estimate}) is that the key parameter $\tau$ should be sufficiently small, $\tau\leq 1$.  This  
  would guarantee that the vacuum transitions would not be strongly suppressed.  The main assumption here is that Aharonov-Bohm (AB) phase coherence can be maintained at sufficient high temperature,  which can drastically decrease parameter $\tau$, see footnote \ref{AB}.

From (\ref{estimate}), one can also set to measure the magnitude of the induced electric dipole moment $\langle d\rangle_{\rm ind}$ which will be generated exclusively as a result of the interaction with the axion field $\theta$. Again, for a system of mm size, the dipole moment according to (\ref{el_moment}), (\ref{m_e}), (\ref{m_e_witten})  is given by
\be
\label{d_estimate}
\langle d\rangle_{\rm ind} \sim 10^{-2}\theta \cdot K_{a\gamma\gamma} \left(\frac{L_3}{1 {\rm mm}}\right) e \cdot \rm cm
\ee
For expected range of $\theta\sim 10^{-18}$, this is about 9 orders of magnitude greater than the current experimental limit of the  electron electric dipole moment (EDM), $|d_e|\lesssim 10^{-29} e \cdot \rm cm$  \cite{Hudson:2011zz,Baron:2013eja}. This large enhancement factor  can be attributed to the coherence of the large number of effective microscopic degrees of freedom participating in generation of the electric dipole moment (\ref{d_estimate})  similar to generation of the magnetic dipole moment (\ref{estimate}) 
as discussed above. The estimate  (\ref{d_estimate}) can be also compared with  the current limit on neutron EDM $|d_{n}| < 10^{-26} e \cdot\rm cm$ \cite{Afach:2015sja}, $\langle d\rangle_{\rm ind}/|d_{n}| \sim 10^{24}\theta\sim 10^{6}$.

One should emphasize that the system is placed in a static uniform magnetic field. In conventional perturbative QED (without topological sectors) even the magnetic moment (\ref{estimate}) cannot be induced  when a ring is placed inside of a uniform static magnetic filed. The generation of a static electric dipole moment 
(\ref{d_estimate}) even a  more puzzling effect within conventional  QED formulated on a trivial topology. The electric dipole moment $\langle d\rangle_{\rm ind} $ could be only generated   due to  a background $\theta$ field, which itself  could be only resulted from the passing nearby axion, or due to fundamental $\theta_{\rm QED}\neq 0$, see more comments on this last  possibility in subsection \ref{QED}.

\subsection{ Emission in the microwave bands }\label{subsec:photonemission}
If the dipole moment (\ref{d_estimate}) is induced due to the passing nearby axion with mass $m_a$, then the system will emit 
 photons  with frequency $\omega=m_a$. Assuming that the system is macroscopically large one can estimate  
 the time-averaged intensity of the dipole radiation for each such unit using the classical expression  
\be
\label{I}
\langle I \rangle \simeq\frac{\omega^4}{12 \pi c^3}\langle d\rangle_{\rm ind}^2\sim \alpha\omega^4 \frac{(L_3 K_{a\gamma\gamma}\theta_m)^2}{12\pi^2c^3},
\ee
where $\theta_m$ is the amplitude of the coherent axion oscillation. 
For a device on the mm scale, again we estimate the average power from a single unit to be on the order of $10^{-13}(K_{a\gamma\gamma}\theta_m)^2 \rm eV^2$, which can be represented in conventional units  as $\langle I \rangle \sim 10^{-16} (K_{a\gamma\gamma}\theta_m)^2$ watt.  Such small intensity is unlikely to be observed even when a large factor $N^6$ for   coherent  emission from $N^3$ dipoles is inserted. 

The effect could potentially be enhanced if a device is specifically designed in a such a way that the splitting of the corresponding quantum levels  exactly coincides with the axion mass. In this case  the quantum  resonance simulated transition is possible which 
may greatly enhance very a low rate of the emission (\ref{I}). Such  computations are well beyond the scope of the present work as they require  some special technique \cite{Yao:2016bps} which  is not yet fully developed.  We leave  the corresponding analysis  for the future studies.

\subsection{Potential Difference and Induced Charge}\label{subsec:potentialdiff}
If we  add 2 plates near the ends of the cylinder,    the induced field $\langle E\rangle_{\rm ind}$ will  induce the electric  charges 
on these plates, which in principle, can be measured. The  magnitude of these charges can be estimated as 
\be
\la Q\ra \sim \frac{e\theta(t)}{2\pi} K_{a\gamma\gamma}. 
\ee
  It is clear that for $\theta$ of interest, $\la Q\ra  \ll e$ and thus the small fractional charges are only to be understood probabilistically. The corresponding potential difference
  can be estimated as 
\be
\label{V}
\la \Delta V\ra \simeq \frac{e\theta K_{a\gamma\gamma}L_3}{2\pi L_1L_2} \sim 10^{-4} K_{a\gamma\gamma}\frac{\theta(t)}{2\pi} (\rm volt).
\ee
To get a bearing on the magnitude of this potential difference, we compare it to the Hall voltage usually on the order of $(10- 100) \rm mV$ in quantum Hall measurements \cite{RevModPhys.58.519,FractionalQHallRev}, which is many orders of magnitude greater that (\ref{V}). The amplification $N^3\sim 10^9$ mentioned above, in principle,  may drastically increase the sensitivity. 

\subsection{Oscillating and Transient Current}\label{subsec:transientcurrent}
As the axion   passes through the detector, the induced dipole and charges on the plate also oscillate, which gives rise to a current in the wire
if we connect the plates. For a superconducting wire, the maximal current one can attain is approximately 

\be
\langle J \rangle \approx \frac{Q c}{L_3}\cdot \left(\frac{v}{c} \right)\sim  \frac{\theta_m K_{a\gamma\gamma}}{2\pi}\cdot  \left(\frac{v}{c} \right) \cdot 10 \rm (nA),
\ee
where $v$ is a typical discharge velocity, which obviously depends on physical properties of  material of a wire. For the superconductor wire we expect that $v\sim c$. 
 The average current can then be measured by a SQUID. Assuming SQUID sensitivity $\sim 10^{-18}\rm T$, in principle each unit can reach sensitivity at $\langle J\rangle \sim 10^{-3} \rm nA$. Therefore, for a clever design of the detector which consists of $N^3\sim 10^9$ such units to amplify the overall current, we are in principle sensitive to $\theta \gtrsim 10^{-12}$.
 
 Based on this estimate, the topological device is in principle competitive with the upper limit of existing and proposed experiments for dark matter axion searches. More importantly, distinct from the direct detection experiments currently known in literature, this configuration is also sensitive to $\theta$ as opposed to $\dot{\theta}$ as a result of the topological features of the system as discussed above. 

Furthermore, one can also use the setup for the measurement of a static $\theta$. For such measurements, we first apply $B_{\rm ext}$ to induce the charges on the plates, and then turn off the applied field and measure the transient discharge current. In next section we elaborate on this option to measure the static $\theta$ using slightly different setups when the topology selects  a specific winding  $|k\rangle$ states
in contrast with specific $|\theta_{\rm eff}\rangle $ states which have been considered so far. The effects will be  still proportional to small parameter $\theta$ 
but some enhancement factors may emerge as we argue below. 

\section{Numerical estimates for $|\kappa\ra$ states  }\label{winding} 
In this section we make some numerical estimates assuming the effective $S^1$ topology is still enforced but in the limit when tunneling and non-trivial winding is suppressed. In this case, $|\kappa\rangle$ states, which are states that correspond to configurations with non-zero flux $\kappa$ threading through space, are good quantum states. Therefore, different from the previous sections, we   perform our analysis in this section for the   $|\kappa\ra$ states rather than for  superposition of the winding states classified 
by parameter $\theta_{\rm eff}$. Physically it implies that we choose uniform magnetic field $B_z$ along $z$ direction 
which selects specific boundary conditions for pure gauge (but topologically nontrivial) vector potential at large distances
such that $\oint_{\Gamma} A_{\mu} dx_{\mu}=2\pi \kappa$, where $\Gamma$ is the path at large distances in $xy$ plane\footnote{The parameter $\kappa$ which classifies our states in the present section is arbitrary real  number. It measures the magnetic physical flux, which   not necessary assumes the  integer values.  It  should not be confused with integer numbers $n, m, k$ which enter the expressions describing the path integral in previous sections, where we sum over all topological sectors to select $|\theta_{\rm eff}\rangle$ state. In present section we select $|\kappa\rangle$ state as the physical state which explains the title of this section \ref{winding}. }.  As we already mentioned in this case   the $\theta$ parameter (not to be confused with $\theta_{\rm eff}$) still remains a physical parameter of the system. 
In such circumstances the electric field (\ref{E}) will be induced along the magnetic field  in the region of  space where the magnetic field is present.   
The topological arguments suggest that the corresponding configurations cannot  ``unwind" as the uniform static magnetic field $B_z$    enforces the system to become effectively two-dimensional, when the $\theta$ parameter is obviously a physical parameter,  similar to analogous analysis  in the well-known  2d Schwinger  model. 

Indeed the $\theta$ term in the   action (\ref{theta}) with fixed $k$ can be rewritten as follows 
 \be
\label{theta1}
S_{\theta}&\sim &  {\theta} e^2\int \dd^4 x ~\vec{E}\cdot\vec{B} =\theta \left[e \int d^2x_{\perp} B_z\right] \cdot\left[e \int dz dt E_z \right]\nonumber\\
&=&2\pi  \kappa ~\theta \cdot \left[e \int dz dt E_z \right].
 \ee
 The expression on the right hand side is still a total divergence, and does not change the equation of motion. In fact, the expression in the brackets is identically the same as the $\theta$ term in 2d Schwinger model,  where it is known to be a physical parameter of the system as a result of nontrivial 
  mapping $\pi_1[U(1)]=\mathbb{Z}$, see e.g. \cite{Cao:2013na} for a short review on $\theta$ term in 2d Schwinger model in the given context. The expression (\ref{theta1}) for the $\theta$ term written in the external background field    shows once again that $\theta$ parameter in 4d Maxwell  theory  becomes  the physical parameter of the system when some conditions are met. In many respects this phenomenon (when the 4d $\theta$ becomes the physically observable parameter of the system)  is very similar to the Witten's effect \cite{Witten:1979ey} when the presence of a monopole enforces the boundary conditions which cannot be 
  ``unwinded" due to the monopole's magnetic topological charge. In such circumstances the $\theta$ term  becomes a physical  parameter of the system (in  monopole's sector), and, in particular, a 
  monopole becomes the dyon with electric charge $\sim \theta$ as we already disused   in section \ref{interpretation}. 
 The role of the magnetic charge (in the Witten's effect) plays    the magnetic flux $\kappa$   in our case. This flux   enforces the boundary conditions (\ref{theta1}) and makes the $\theta_{\rm QED}$ to become an observable parameter in the sector with non-vanishing magnetic flux.  
    
The   discussions of  the present  section are  devoted to the physically observable effects due to $\theta$ parameter (\ref{theta1}) in the given $|\kappa\ra $ sector. In this case   we do not have any other requirements (such as small $\tau$) except that AB coherence phase must be  maintained. This  opens up a new perspective for the axion search experiments due to two reasons. First, the effect is sensitive to the static   axion field   even  without tunnelling suppression factor $ \sim \exp(-1/e^2)$ which always accompanied  all our formulae in the previous section. 
This is because we study the system in the $k$ sector with non vanishing $\kappa$ flux, similar to the Witten's effect with non vanishing 
monopole's charge. In both cases the effect is proportional to $\theta$ without  $  \exp(-1/e^2)$ suppression because the object of studies is not the vacuum, but the heavy $k$ sector. 

Secondly,  the effect for the induced electric field (\ref{E}) and  related formulae for the potential difference $\la \Delta V \ra$, the induced charges $\la Q\ra$ and induced currents 
$\la J\ra$ can be drastically  enhanced due to the additional parameter $B^z_{\rm ext}$ which could be in the range of Tesla rather than in a fraction of  Gauss, see next subsection \ref{enhancement} with detail estimates.  
This is because   the overall expression is approximated by a linear dependence on the external field. Such a  linear dependence on the external field  for the   $|\kappa\ra$ states can be easily understood from
expression (\ref{theta1}) where $\theta$ parameter  always enters in combination with (non-fluctuating)  external field $B_z$, 
which eventually drastically enhances all the effects related to $\theta$.
\exclude{Nevertheless, the periodicity $\theta_{\rm eff}\rightarrow  (\theta_{\rm eff}+2\pi n)$ is ultimately enforced by the infinite sum of exponentially suppressed terms in (\ref{E_ind}). In the Poisson resummed expression, the contributions are only coming from very large $n$, which explains the linear dependence of the effects on $B_z$ for leading order contributions.} 

For completeness of the presentation we also briefly discuss in section \ref{dual} the dual picture when the magnetic field is induced in the background of the electric field. We    emphasize  in section  \ref{QED} that  our studies on static $\theta$ due to the DM axion passing the detector can be equally apply to constraint the fundamental parameter of QED, the $\theta_{\rm QED}$ which becomes a physically observable  parameter of the system  when the theory is formulated on a nontrivial manifold, or it  is placed  into the background field which itself enforces a  nontrivial topology  as   discussed   above.

\subsection{Potential Difference, Induced Charge and Induced Current with $B_{\rm ext}$}\label{enhancement}
From (\ref{E}), we arrive at the following expression for the induced electric field in the presence of $\theta\neq 0$.
\be
\label{E1}
\la E\ra_{\rm ind} =- \frac{\theta K_{a\gamma\gamma}\alpha}{\pi} B_{\rm ext}.
\ee
If we place the plates at the ends of the cylinder, the induced field  $\la E\ra_{\rm ind}$ will induce the charge on the plates 
similar to our discussions in the previous section. The magnitude of the charges can be estimated as follows
\be
\label{Q1} 
\la Q\ra \sim \frac{e\theta(t)}{2\pi} K_{a\gamma\gamma}\cdot\left[\frac{e B_{\rm ext} L_1L_2}{2\pi}\right]. 
\ee
 The difference, in comparison with the previous discussions, is that the external magnetic field $B_{\rm ext}$ could be quite large
 which makes the effect much stronger. This charge separation effect due to $\theta\neq 0$ generates the potential difference, 
 which can be estimated as follows 
 \be
\label{V1}
\la \Delta V \ra &\simeq& \frac{e\theta K_{a\gamma\gamma}L_3}{2\pi L_1L_2} 
\cdot\left[\frac{e B_{\rm ext} L_1L_2}{2\pi}\right]\\
&\sim & 0.2 K_{a\gamma\gamma} \theta \cdot \left(\frac{L_3}{\rm mm}\right) 
\cdot \left(\frac{B_{\rm ext}}{\rm Gauss}\right)  (\rm volt).\nonumber
\ee
If we place the system into the background  of the strong external field 
 $B_{\rm ext}\sim 1$T  and  assume the   $N^3\sim 10^9$ amplification mentioned above,  along with  high sensitivity to measure the potential difference on the level $10 \rm mV$ (which is typical for  the quantum Hall measurements \cite{RevModPhys.58.519,FractionalQHallRev}), then one can push the sensitivity for the static $\theta$ to the level $\theta \sim 10^{-14}$. We emphasize that such a measurements are sensitive to the static $\theta$ regardless of its nature: whether it is the fundamental constant of QED, or it is generated by the coherent DM axion passing through detector. 
 
 If we connect two plates with a   wire, then the induced current can be estimated as follows   
 \be
 \label{J1}
\la J \ra &\sim& \frac{\la Q\ra c }{L_3}\left(\frac{v}{c}\right)  \\  &\approx &10^{-6} \cdot \left(\frac{\theta K_{a\gamma\gamma} }{10^{-14}}\right)\cdot \left(\frac{L_1L_2/L_3}{\rm mm}\right) \cdot \left(\frac{B_{\rm ext}}{\rm  1 T}\right)\cdot \left(\frac{v}{c}\right) {\rm nA}\nonumber
\ee
where $v$ is a typical discharge velocity, which obviously depends on physical properties of  material of a wire. For the superconducting wire we expect that the discharge velocity is close to maximum possible  $v\sim c$.  A single device obviously cannot 
produce a measurable  effect for   small $\theta\simeq 10^{-14}$. However, as we already mentioned, the potential difference (\ref{V1}) for  such small  $\theta\simeq 10^{-14}$  can be in principle measured  with  the   $N^3\sim 10^9$ amplification. A similar amplification  produces the  effect  for the current  $ \la J \ra  $ on the level $10^3 ~{\rm nA}$ which is the same order of magnitude as   in the usual experiments that measure  persistent currents
\cite{Buttiker:1983,Chandrasekharetal:1991, Shanks:2011}. 
The challenge, of course,  will be in maintaining Aharonov-Bohm-like coherence in the presence of a strong magnetic field using superconducting wire  for the given system size.   

To reiterate the basic result of this subsection: the primary phenomenon of  this system is  the generation of the induced electric field (\ref{E1})  in the background of the external magnetic field and in the presence of non-vanishing  $\theta$. One can, in principle, observe this   small electric field for $\theta\simeq 10^{-14}$ by  measuring the induced charges   (\ref{Q1}), the potential difference (\ref{V1}), or the discharge current (\ref{J1}), all of which obviously represent the secondary effects which follow from   (\ref{E1}). 

\subsection{Dual Picture with  electric  external field  $E_{\rm ext}$}\label{dual}
In this subsection we want to elaborate on  the dual picture of the same phenomenon when the external magnetic field $B_{\rm ext}$ is replaced by external electric field $E_{\rm ext}$. In this case, formula (\ref{dual_indfield}) suggests the magnetic field is induced.   
In physical notations  using the    Minkowski  signature this formula reads
\be
\label{B1}
\la B\ra_{\rm ind} = \frac{\theta K_{a\gamma\gamma}\alpha}{\pi} E_{\rm ext}, 
\ee
which  represents, in all respects,  the dual expression for  (\ref{E1}) when electric and magnetic fields  exchange their roles: the external field becomes the  induced field, and vice versa. The significance of this formula is that the induced magnetic field can be measured with very high accuracy  
 by making use of the existing infrastructure in the usual experiments that measure  persistent currents from Aharonov-Bohm coherence 
 \cite{Buttiker:1983,Chandrasekharetal:1991, Shanks:2011}. 
 
 Instead of applying a magnetic field through the ring, however, one can instead apply an external electric field in the same direction. 
As  eq. (\ref{B1}) states, the   magnetic flux will be induced along the principal axis of the small cylinder/ring. The induced magnetic flux can be understood \cite{Zhitnitsky:2015fpa} as coming from the surface persistent current $\la J \ra$ which can be estimated as follows 
\be
\label{J2}
\la J\ra&\simeq& \frac{\theta K_{a\gamma\gamma}\alpha}{\pi} L_3 E_{\rm ext}\\ &\sim& 10^{-6}\cdot\left(\frac{\theta K_{a\gamma\gamma}}{10^{-14}}\right)\cdot  \left(\frac{E_{\rm ext}}{10^5\frac{\rm V}{\rm cm}}\right)\cdot  \left(\frac{L_3}{\rm mm}\right) {\rm nA}. \nonumber
\ee
This estimate shows that the magnitude of the current is the same order of magnitude as (\ref{J1}) for the discharge current  for   small $\theta\simeq 10^{-14}$. However, as before, the current (\ref{J2})    can be in principle measured  with  the   $N^3\sim 10^9$ amplification. A huge advantage of this specific design  is that the physics of the persistent currents is well understood.
For zero  external magnetic field the persistent currents are obviously vanish. The currents may only be generated  if the $\theta$ parameter does not vanish. It could only happen if  the DM axions $\theta(x)$ are passing through detector, or due to the fundamental $\theta_{\rm QED}\neq 0$, see next subsection with  some comments on this possibility.  We emphasize again that the effect (\ref{J2}) is proportional to the static $\theta$, rather than $\partial_{\mu}\theta$  as explained in the details at the very beginning of this section \ref{winding}.

\subsection{Measurement of fundamental $\theta_{\rm QED}$\label{QED}}
In section \ref{numerics} we studied a number of effects which are sensitive to constant $\theta$.
All  these effects are due to the tunnelling transitions between states that are topologically distinct but related by (large) gauge transformations. Therefore, all the effects are  formally suppressed by a factor  $\exp(-1/e^2)$,  though numerically the suppression could be  quite mild as long as parameter $\tau \sim1$. On other hand, in subsections (\ref{enhancement}) and (\ref{dual}) we studied a number of effects which are also sensitive to constant $\theta$. However, in that case the effects considered are in the suppressed tunnelling limit. As explained at the very beginning of  section \ref{winding} the optimized effects are proportional to $\theta$    without  $  \exp(-1/e^2)$ suppression  because the system belongs to the  sector with non-vanishing magnetic flux, similar to the Witten's effect \cite{Witten:1979ey} with non-vanishing 
monopole's charge. In both cases the effect is proportional to $\theta$ (and not to $\partial_{\mu}\theta$ ) because   the  systems belong to the heavy $\kappa$ sectors, rather than to unique vacuum sector where  factor $\exp(-1/e^2)$ unavoidably   emerges in  the Maxwell system   for non-simply connected manifolds with nontrivial $\pi_1[U(1)]$. 

The numerical estimates in subsections (\ref{enhancement}) and (\ref{dual}) are quite promising as they show   a number of potentially  enhanced  factors which in principle can drastically increase the sensitivity to constraint (or observe) very tiny $\theta \gtrsim  10^{-14}$.
The same analysis    can also be used to   put upper limit on the fundamental constant of $\theta_{\rm QED}$, a parameter that has not been measured to the best of our knowledge. While of $\theta_{\rm QED}$ does not produce any physically measurable effects for QED with trivial topology, or in vacuum,  we expect the proposed Aharonov-Bohm-type configuration discussed in sections \ref{numerics} and \ref{winding} to be sensitive to such a parameter which is normally ``undetectable'' in a typical scattering experiment based on perturbative analysis of QED. 

The limits imposed by dipole moment, charge, potential and transient current will be the same as the limit set for  the coherent DM axion $\theta$. For a detector consists of $N^3\sim 10^9$ such units, $\theta_{\rm QED}$ one should in principle be able to exclude $\theta_{\rm QED}\gtrsim  10^{-14}$, depending on the specific realizable experimental setup.

We conclude this section with the following remarks related to the previous studies.  First, the fact that  $\theta_{\rm QED}$ becomes a physically observable parameter when  the theory is formulated on a nontrivial manifold 
has been known  since  \cite{Witten:1995gf,Verlinde:1995mz,Olive:2000yy}. Our original contribution into this field is  represented by 
a number of  explicit formulae derived for simple geometries when nontrivial    topological features
of the system   manifest themselves. Precisely these simple and explicit formulae  allow us to produce a number of numerical estimates
of the $\theta$ related effects. These explicit estimates could be the important   elements relevant for  the novel types of the axion search experiments 
because the conventional searches for the dark matter axions are mostly based on effects  when the axion field enters via the derivatives $\sim \partial_{\mu}\theta(x)$.

Secondly, the possible physical effects from $\theta_{\rm QED}$ has also been previously discussed \cite{Hsu:2010jm,Hsu:2011sx} in the   spirit of the present work. Here we have proposed a more detailed experimental setup with estimations on sensitivity that is in principle experimentally accessible. More importantly, we explicitly computed both type of the effects, with and without $\exp(-1/e^2)$ suppression to emphasize the role of topology, irrespectively   whether it is enforced by external field in  $\kappa$ sectors as in section \ref{winding}, or by a  design of a  non simply connected spatial manifolds as in section \ref{numerics}.

            \section{Conclusion}\label{conclusion}
            The main goals of the present paper   can be summarized on three  distinct but related items.
            
   1.         First of all, we studied a  fundamental,  time independent $\theta_{\rm QED}$ term in QED which is known to become  a physical observable when  some conditions are met  \cite{Witten:1995gf,Verlinde:1995mz,Olive:2000yy}.  We produced a number of explicit computations for simple geometries where  $\theta_{\rm QED}$ related effects can be easily understood.   This should be   contrasted with conventional viewpoint that $\theta_{\rm QED}$  is not a physically relevant parameter in abelian gauge theories, and must enter the  observables in form of $\partial_{\mu}(\theta_{\rm QED})$.
   
    A deep reason why $\theta_{\rm QED}$ in vacuum becomes a physical parameter when the Maxwell  system is defined on non-simply connected manifold is due to the  emergence of the so-called Gribov's ambiguities (when the gauge cannot be completely fixed), which is well known phenomenon in non-abelian gauge theories,  see footnote \ref{Gribov} for comments and  references. The $\theta_{\rm QED}$ also becomes a physically observable parameter in the sectors with non vanishing  magnetic flux  $\kappa$  as explained in section \ref{winding}, and more specifically, in subsection \ref{QED}. In this case the effect is very much the same as the Witten's effect 
   when the $\theta$ becomes a physically observable parameter in the monopole's sectors. 
      
   We suggest a specific design for a tabletop experiment  which, in principle,  can constrain the fundamental $\theta_{\rm QED}$ on the level $\theta_{\rm QED}\sim 10^{-14}$ .  This constraint should be treated as an independent  from known constraint on a physically  distinct  parameter $\theta_{\rm QCD}\leq 10^{-10}$.  
 
   2. A related, but distinct, goal is the application of our findings  to the axion search experiments, where $\theta(x)$ describes the DM axion with typical frequency 
            $\omega=m_a$ passing through a detector.  This part of the paper  is motivated by a number of  ongoing axion search experiments which may finally unlock the nature of the dark matter.    The novel element which was not previously fully explored is that 
            the observable effects   may depend directly on $\theta$ rather than on $\partial_{\mu}\theta$ which normally enters the conventional formulae devoted to the DM axion search experiments. The topological arguments play the crucial difference, similar to item 1 above.  Additionally, the proposed experiment can in principle probe the entire open mass window, in contrast with the conventional resonant cavity techniques.
            
            3. The final  goal   of this work is to advocate  an idea that there is a novel type of vacuum energy, the Topological Casimir Effect (TCE) which cannot be formulated in terms of propagating degrees of freedom (the photons in the Maxwell theory). This new type of energy is highly sensitive to the $\theta (t)$ parameter.
            Therefore, there is a real chance to ascertain the existence of such effects using specifically designed    instruments with  high  precision,  as discussed in this paper.  The work in this direction  might shed some light on the nature of the dark energy. In fact, the original papers   \cite{Cao:2013na,Cao:2015uza} on TCE were  mostly  motivated  by the idea to imitate  this new type of vacuum energy in a tabletop experiment. 
             
            In the context of QCD,  this type of the   vacuum energy cannot be   associated with   any physical propagating degrees of freedom, analogous to TCE in  the present work in Maxwell system.  This motivates the proposal  in  \cite{Zhitnitsky:2013pna, Zhitnitsky:2015dia}  that the  observed dark  energy in  the Universe may have, in fact,  precisely such a non-dispersive  nature\footnote{This novel type  of vacuum energy which can not be expressed in terms of propagating degrees of freedom has been extensively studied in QCD lattice simulations, see \cite{Zhitnitsky:2013pna} and references therein on the original lattice results.}. The proposal where an
             extra energy\footnote{There are two instances in the evolution of the Universe when the vacuum energy plays a crucial  role.
The first instance   is identified with  the inflationary epoch  when the Hubble constant $H$ was almost constant, which corresponds to the de Sitter type behaviour $a(t)\sim \exp(Ht)$ with exponential growth of the size $a(t)$ of the Universe. The  second instance where the vacuum energy plays a dominant role  corresponds to the present epoch when the vacuum energy is identified with the so-called dark energy $\rho_{DE}$ which constitutes almost $70\%$ of the critical density. In the proposal  \cite{Zhitnitsky:2013pna,Zhitnitsky:2015dia}  the vacuum energy density can be estimated as $\rho_{DE}\sim H\Lambda^3_{QCD}\sim (10^{-4}{\rm  eV})^4$, which is amazingly  close to the observed value.}  cannot be associated with any propagating particles is aimed to provide an approach that is different from a commonly accepted paradigm that the extra vacuum energy in the Universe is always associated with the potential of some  propagating degree of freedom, 
            such as inflaton, see original papers \cite{inflation1,inflation2}, and reviews \cite{linde, mukhanov}.


 \section*{Acknowledgements} 
 
    We would like to thank Sean Carroll, Stephen Hsu, Zitao Wang, and Mark Wise for helpful discussions. 
    One of us (AZ) is thankful to Cliff Burgess, Glenn Strakman, Mark Trodden and other participants of  the MITP program ``Quantum Vacuum and Gravitation", Mainz,  March, 2017, 
    for very useful discussions related to the subject of the present work.   AZ is also thankful to Giovanni Cantatore for long and very optimistic  discussions on feasibility  to experimentally test the TCE and its sensitivity to $\theta$.  
    This research was supported in part by the Natural Sciences and Engineering Research Council of Canada and by the Walter Burke Institute for Theoretical Physics at Caltech, by DOE grant DE-SC0011632, by the Foundational Questions Institute, by the Gordon and Betty Moore Foundation through Grant 776 to the Caltech Moore Center for Theoretical Cosmology and Physics, and by the John Simon Guggenheim Memorial Foundation. 
    

 \bibliography{Axion-TCE}
\bibliographystyle{JHEP}

\end{document}